\documentclass[onecolumn,showpacs,preprintnumbers,amsmath,amssymb,prb]{revtex4}

\usepackage{latexsym}
\usepackage{amssymb, amsmath}

\usepackage{graphicx}
\usepackage{dcolumn}
\usepackage{bm}

\usepackage[colorlinks=true, pdfstartview=FitV, linkcolor=red, citecolor=blue, urlcolor=blue]{hyperref}

\newcommand{\dbb}{de Broglie-Bohm}
\newcommand{\Eq}{Equation}
\newcommand{\Eqs}{Equations}
\newcommand{\eg}{{\it e.g.}}

\begin{document}



\title{Local  de Broglie-Bohm  Trajectories from Entangled Wavefunctions}

\author{Michael Clover}
\email{michael.r.clover@saic.com}
\affiliation{%
Science Applications International Corporation\\
San Diego, CA \\
}

\date{\today}
\begin{abstract}
{We present a local interpretation of what is usually considered to be a nonlocal  \dbb~trajectory prescription for an entangled singlet state of massive particles.  After reviewing various meanings of the term ``nonlocal'', we show that by using appropriately retarded wavefunctions (i.e., the locality loophole) this local model  can violate Bell's inequality, without making any appeal to detector inefficiencies.  

We analyze a possible  experimental configuration appropriate to massive two-particle singlet wavefunctions and find that as long as the particles are not ultra-relativistic, a locality loophole exists and  Dirac wave(s) can propagate from Alice or Bob's changing magnetic field, through space, to the other detector, arriving before the particle and thereby allowing a local interpretation to the 2-particle de Broglie-Bohm trajectories. 

 We also propose a physical effect due to changing magnetic fields in a Stern-Gerlach EPR setup that will throw away events and create a detector loophole in otherwise perfectly efficient detectors, an effect that is only significant for near-luminal particles that might otherwise close the locality loophole.   }
\end{abstract}

\maketitle
\noindent

\section{Introduction}

In the EPR literature, it is usually considered that the only local models that can explain EPR experiments are contrived intellectual diversions~\cite{mrc1,PH02,TH02}, bearing no relation to physical reality, and that additionally  require some ``unfair'' sampling mechanism to generate the correct experimental results~\cite{GZ99,GG99,SF02,L99,L03}.  Somewhat less contrived, Accardi~\cite{Ac06} has proposed a chameleon model that is claimed to be local and violate Bell's inequalities, and Christian~\cite{C0703179} had proposed a local model based on properties of a Clifford algebra.  With the exception of the last two, most local hidden variable models assume a ``detector'' loophole.  

It also seems to be conceded that  -- for a single  massive particle\footnote{We will confine our discussion to massive particles, for which there is a well-defined de Broglie-Bohm interpretation based on the first-quantized Schr\"odinger and Dirac wavefunctions.
} 
--  the de Broglie-Bohm interpretation of quantum mechanics is local, in that the gradients of the wave function at the particle position provide a quantum field that exerts a force on a deterministic, otherwise classical, particle. There seems to be general  agreement  that when it comes to two entangled particles, the same \dbb~interpretation is manifestly nonlocal~\cite{B52a,B66}.  It is with this argument that we take  issue, and  propose to regard the two-particle \dbb~relations~\cite{DP02} as a \emph{local}  model  -- which we claim can be done as long as time-retarded wavefunctions are used. We motivate this by an example:  an analysis of Hooke's Law and its apparent nonlocality.  

 We will show that our local \dbb~model can violate Bell's inequality as long as it can use the ``locality'' loophole.     Since QM (Dirac) waves move at a phase velocity of $c$, and since massive particles move much slower, this loophole appears to be big enough  for all conceivable experiments involving massive particles. We will also show that there exists a physical mechanism that could operate on these massive particles to provide a ``detection'' loophole, which we will see is not significant unless the particles are moving at ultra-relativistic speeds -- but for such Stern-Gerlach experiments, means that one could still measure apparent violations of Bell's inequality with a local mechanism.

\section{How Non-local is de Broglie-Bohm?}

In the single particle case, the \dbb ~velocity law takes the form
\begin{eqnarray}
\vec{v}(x,t)  &=& -\frac{i\hbar}{2m}\left( \frac{ \psi^* (x,t)\nabla \psi(x,t) - (\nabla \psi(x,t))^* \psi(x,t)}{\psi^*(x,t) \psi(x,t)}\right)  =  \frac{\hbar}{m} \Im \left( 
\frac{\nabla \psi(x,t)}{\psi(x,t)} \right) , \label{eq:vdbbs}
\end{eqnarray}
where $\psi(x,t)$ is the solution to the Schr\"odinger equation.  This is a non-linear, but not obviously non-local, prescription for the velocity of the particle.
  If a boundary condition changes far away (e.g., closing a slit at $x=-L$ at time $t=0$), and the effect is felt at $x=+L$ simultaneously, we will claim that the nonlocality is a result of using the  non-relativistic Schr\"odinger equation; using the relativistic Dirac equation means that the wavefunction at $x=+L$ will not change until after the time $\Delta t= 2L/c$, when the ``Dirac wave'' has propagated to that position (the velocity prescription for Dirac wavefunctions~\cite{H99} is similar to that of \dbb's for Schr\"odinger wavefunctions: $dx^i/dt = c j^i/j^0$ where $j^{\mu} = c \overline{\psi} \gamma^{\mu} \psi$).    Thus, for a single particle, we claim that \dbb~interpretation is not {\em essentially} non-local.
  
   In the case of two particles, we have 
  \begin{eqnarray}
\vec{v}_1( t )  & \sim &   \frac{\hbar}{m} \Im \left( 
\frac{\nabla_1 \psi(x_1,x_2,t)}{\psi(x_1,x_2,t)} \right) = v_1(x_1,x_2,t) \ne v(x_1,t)\ , \notag \\
\vec{v}_2( t )  & \sim &   \frac{\hbar}{m} \Im \left( 
\frac{\nabla_2 \psi(x_1,x_2,t)}{\psi(x_1,x_2,t)} \right) = v_2(x_1,x_2,t) \ne v(x_2,t)\ , \nonumber 
\end{eqnarray}
for the velocity of each particle, and see that, formally at least, neither particle's velocity is strictly a function of it's own position, as one might expect for classical particles in local potentials.  This dependence is taken as {\it a priori} evidence of nonlocality of the \dbb~interpretation~\cite{Sq93,holland}.  We shall argue below  that this dependence is misinterpreted.

Bohm's justification of the physicality of nonlocality ({\it i.e.}, the transmission of forces at infinite speed) was to argue that it is a ``non-signaling'' non-locality, since his interpretation had been shown to lead\cite[p. 187]{B52a} to ``precisely the same predictions for all physical processes as are obtained from the usual interpretation (which is known to be consistent with relativity).''  In other words, by taking the ensemble average of lots of non-local results he gets a local result.  

We would prefer the grammatically simpler ``locality'' to the double negative of a non-signaling non-locality. As far as we can see, the only reason most physicists don't agree, is that Bell has said that certain experimental measurements are evidence of nonlocality.  We have argued elsewhere~\cite{mrc3,mrc4} that Bell's derivation assumed a certain non-contextuality that was too strict, and that a weaker limit could be derived for hidden variables that are ``locally'' contextual (by local contextuality we mean hidden variables that prevent Alice from measuring $a'$ at the same time she measures $a$).  This paper presents an example of a local hidden variable theory that is also  ``remotely'' contextual, in that Alice's result depends on her causal knowledge of Bob's detector setting ({\it i.e.}, at a retarded time, $A(a(t),b(t-D/c),\lambda(t))$).

\subsection{How non-local is Hooke's Law?\label{secn:hooke}}

Consideration of a wider relevant context prompts one to ask   ``How non-local is Hooke's Law?'', because in that case we also have a force law that looks ``manifestly'' nonlocal:
\begin{subequations}\label{eq:two}
\begin{align}
m_1 \ddot{x}_1 &= -k (x_1 - x_2) \ , \label{eq:2a} \\
m_2 \ddot{x}_2 &= +k (x_1 - x_2) \ . \label{eq:2b}
\end{align}
\end{subequations}
We are not aware of any textbooks that  use this as an example of non-locality (springs being an archetypical example of local, causal wave propagation), and most physicists would expect that even if $m_2$ were suddenly fixed (in \Eq~\eqref{eq:2a}), that $m_1$ wouldn't react to that event until a sound wave had passed down the spring from $m_2$ to $m_1$.  That being the case, perhaps $x_2$ should be retarded in the first equation and $x_1$ in the second; unless it is invalid to use these equations in that context.

The \dbb~interpretation has been modified in precisely that manner~\cite{Sq93,Sq95,MS95} --  to use time-retarded positions in a time-independent wavefunction -- in order to force a local interpretation out of the velocity prescriptions.  This means, of course, that if the original velocity prescription  came from nonlocal non-signaling interpretations of the Dirac equation, that the new interpretation would be the solution to something that isn't quite the Dirac equation anymore, and sooner or later, for some measurement this new re-interpretation  will have to give a theoretical prediction  different from the Dirac equation (thereby falsifying this retarded solution).

If the positions in Hooke's law are retarded, we would then have
\begin{eqnarray}
m_1 \ddot{x}_1(t) &=& -k (x_1(t) - x_2(t-\tau)) \ , \nonumber \\
m_2 \ddot{x}_2(t) &=& +k (x_1(t-\tau) - x_2(t)) \ , \nonumber
\end{eqnarray}
where $\tau = \Delta x/c_s$, and $\Delta x$ is some measure of the separation.  One can expand these new expressions about $t$ for small $\tau$, leading to
\begin{eqnarray}
m_1 \ddot{x}_1(t) &=& -k (x_1(t) - x_2(t))-\tau k \dot{x}_2(t) \ , \nonumber \\
m_2 \ddot{x}_2(t) &=& +k (x_1(t) - x_2(t)) -\tau k \dot{x}_1(t) \ , \nonumber
\end{eqnarray}
which still has the force on the first particle depending on the instantaneous position of the second particle, and  {\it vice versa}, but now a damping force ($\propto c^{-1}$) depends on the instantaneous velocity of the other particle as well.  It is possible that this {\it reductio ad infinitum} might converge short of {\it absurdum}, but we would suggest that retarding something that is not nonlocal in the first place will just make the local equations wrong.

Our preferred interpretation of this case is to take seriously the notion that Equations~\eqref{eq:two} are an {\em idealization only appropriate to an isolated system}.  If the system is isolated, then no one can grab one particle to initiate a nonlocal signal in the first place, and if they do, then the system is no longer isolated, the equations no longer apply, and there is no issue of whether invalid equations can signal or not.

 Nonetheless, within the idealization, there is a way for $m_2$ in equation~\eqref{eq:2b} to show indifference to $m_1$'s fate for at least a while, and that would be if $m_2$ took seriously the notion that it was part of an isolated system.  Suppose $m_2$ ``knows'' not only his current location, but has ``calculated'' (with \Eq~\eqref{eq:2a}) where the coupled and equally {\em isolated} $m_1$ should be and thus what acceleration $m_2$ should be experiencing (similarly, $m_1$ calculates  $m_2$'s location in order to update her acceleration, in both cases using initial conditions learnt from earlier causal contact).  In that case, $m_2$ will continue to oscillate regardless of whether something grabs $m_1$ at a particular time or not.  Of course, $m_2$ will continue to oscillate even after $\Delta x/c_s$, but now we know that our (local) mathematical model is in a regime it is not designed to handle (and we    can predict when any discrepancies should occur); we would have to go beyond two isolated mass points to the continuum model of Navier-Stokes equations in order to have enough sophistication to propagate sound waves from changing boundary conditions at one location  to other locations.  
 
 Note that we can also transform the equations into another form that suggests more locality and less computational capabilities for mass points:
 \begin{eqnarray}
m_1 \ddot{x}_1(t) &=& -k (x_1(t) - x_2(t)) \ , \nonumber \\
 &=& -2k (x_1(t) - X_{cm}) \ .  \notag 
 \end{eqnarray}
As long as the center of mass of the isolated system behaves predictably ($X_{cm}(t) = X_0$, which can always be done after a Gallilean boost if $\dot{X}_0 \ne 0$), the only non-parametric variable in the force on $m_1$ is the position of $m_1$ in the ``force-field'' of the center of mass, {\it i.e.}, $F_1 = F(k,X_0; x_1)$.

In this way we see that an isolated system of particles can be such that one particle can depend on the instantaneous position of another, without requiring anything but a local, causal force field at each point (which effectively calculates $m_1$'s position for $m_2$ and {\it vice versa}).   Without the concept of a force-field, Newton was flummoxed less by the nonlocality than the action at a distance;  with the concept of a quantum potential and its field extending through the aether,  there is no problem of $m_1$ acting on $m_2$ from a distance, although the question of whether it does it instantaneously or not requires us to determine what an ``isolated'' system is.

  We will suggest that  in an EPR experiment, by knowing $x_1(0)$, $x_2(0)$, and $\Psi$, Alice can calculate both  $x_1$ and $x_2$ given the \dbb~velocity prescriptions, and similarly for Bob.
  For Hooke's law, the differential equations describing the motion are only correct as long as the system is undisturbed -- any attempt to grab one of the mass points destroys the isolation  and changes the boundary conditions of an underlying continuum model; for an EPR experiment, the Schrodinger wavefunctions provide valid trajectories as long as the boundary conditions (Alice and Bob's magnet settings) are unchanged -- otherwise, we have to wait for the Dirac waves to re-equilibrate the new settings.  We remind the reader that the usual method of calculating \dbb~trajectories is to solve $i\hbar \dot{\psi} = H\psi$ in order to propagate a wave packet from its initial position to the final position, so that the derivatives of that wave-packet wavefunction can provide the pilot instructions to the individual point particles.    If the boundary conditions change during that time, it is clear that a more fundamental theory must be invoked --  for example the Dirac equation, which will propagate changes to the wavefunction at the speed of light.  Even so, this still leaves the $x_1-x_2$ type functional dependence, which can be worked around in the same way that Hooke's $m_1$ and $m_2$ performed redundant calculations, by assuming that individual particles know the (fixed) position of the center of mass.  If a more fundamental approach is needed, one can then resort to second quantization and QED, the continuum analogs to Navier-Stokes.

As with springs,  for two quantum-mechanical particles one can recast a Schr\"odinger equation using individual  coordinates, $\psi(x_1,x_2)$, into one using center-of-mass and relative coordinates,  $\Psi(X,x)$, instead.  In the Stern-Gerlach case,  there are some practical advantages working with two coordinate systems that are not aligned with respect to each other (where Bob's $\hat{x}_2  \parallel   \hat{B}_{bob}$ and  Alice's $\hat{x}_2  \parallel  \hat{B}_{alice}$,   but $ \hat{B}_{bob} \nparallel  \hat{B}_{alice}$), and since the singlet state's relative coordinate wavefunction still entangles spin up and down, there is no advantage to using relative coordinates for this problem.   ({\it N.B.}: when Alice and Bob's magnets are at the same angle, so that the two $\hat{x}$'s are parallel, the relative coordinate formulation does allow one to see that the \dbb~velocity law factors into a single function of $\vec{x}$, independent of $\vec{X}$, shown in \Eq~\eqref{eq:dpmatch} below. In other words, for aligned magnet settings, the \dbb~velocity law takes on the same dependence on relative coordinates as a Hooke or Coulomb force law, which we have seen is local as long as the system is isolated.)

\subsection{ Entanglement and {\it a priori} Nonlocality\label{secn:entangl}}

Rice~\cite{Rice97} presents a geometric argument that the \dbb~interpretation must be nonlocal, arguing, essentially, that an entangled wavefunction has nonlocality built into it despite (or in spite of) the local behavior of a single wavefunction.
Consider the case of two non-interacting particles, such that  $H_1 \psi_1 = i\hbar \dot{\psi_1}$ and $H_2 \psi_2 = i\hbar \dot{\psi_2}$. A basis-state wavefunction for the two-particle system would have the form $\psi_1(x_1,t)\psi_2(x_2,t)$, and an entangled, singlet state wavefunction can  be written as $\psi_1(x_1,t)\psi_2(x_2,t)-\psi_1(x_2,t)\psi_2(x_1,t)$.  It is this form that Rice claims is nonlocal:  start by assuming that Bob makes some change to his boundary conditions at time $t=0$.   Alice (who we assume is at a distance D from Bob), in order to apply the \dbb~method to describe the position of her particle ($m_1$), determines that her $v_1$ is proportional to $\psi_2(x_2,t) \nabla_1 \psi_1(x_1,t)-\psi_1(x_2,t)\nabla_1\psi_2(x_1,t)$;  if the time $t$ is less than $D/c$, then no causal signal from Bob ({\it e.g.}, changes in his Dirac wavefunction) should have reached Alice.  Nonetheless, the velocity that she calculates has a (first) term proportional to $\psi_2(x_2,t)$ --  her velocity depends on  Bob's wavefunction ($\psi_2$) after his change to the boundary condition  but before Alice can possibly know about it in any causal manner, even if Alice has been keeping track of where Bob's $m_2$ is  or should be currently located.  In other words, it is not the $\psi(x_2)$ dependence that is non-local, it is the $\psi_2(t)$ that is non-local regardless of where $x_2$ is at.  Whether these wavefunctions are solutions to Dirac  or Schr\"odinger equations, it is this {\em form} of the entangled wavefunction that Rice claims is {\it a priori} nonlocal.

In regular quantum mechanics, the only use made of such entangled wavefunctions (or any wavefunction) is to calculate matrix elements, and the issue of nonlocality does not arise in such a case: one integrates over all $dx_2$ at a time $t$, resulting in something that is only a function of $x_1$, which in turn is integrated over $dx_1$ to produce the number that is compared to some experimental measurement.  (Note that during the integration over $dx_2$ and $dx_1$, it is assumed that the wavefunction is static, corresponding exactly to an experiment done with  detectors in fixed positions.  If Alice and Bob are rapidly changing their settings, then the quantum-mechanical averages need to be done with more care than has yet been shown by anyone in the EPR community.) Only the \dbb~interpretation makes use of the ``pointwise'' values of the wavefunction in its velocity prescriptions, and so only the \dbb~interpretation suddenly confronts this peculiar form of nonlocality.  

There are three possible solutions to this problem.  The historical approach (Rice's) is to assume that Alice is indeed godlike (or at least nonlocally omniscient), and instantaneously knows what Bob's wavefunction has become, so that she can  apply her nonlocal velocity algorithm to determine $v_1$, and we again have a ``non-signaling'' nonlocality.  Another approach (Squires') is to retard the position of the opposite particle in the (presumably still nonlocal) time dependent wavefunction.\footnote{Squires' work~\cite{Sq93,Sq95,MS95} retarded positions in time {\em independent} wavefunctions.} 

The alternative which we propose is to insist that Alice can only make use of locally available information, and that while she can use Bob's wavefunction to calculate her velocity, it must be an appropriately retarded wavefunction -- the one that Bob would have had if he had done nothing in the most recent $D/c$ time interval.  Unlike the classical Alice, who only had to calculate $x_1$ and $x_2$ from Hooke's law, the quantum Alice not only has to calculate $x_1$ and $x_2$, she has to calculate both $\psi_1$ and $\psi_2$ using only her causal information about all boundary conditions (ditto for quantum Bob). 

 It is hard to pick a good notation with which to write this; Alice will calculate a contemporaneous wavefunction, $\psi_2(x_2,t)$, to describe the velocity of Bob's particle at the current time and non-retarded position, but this PDE will be solved using boundary conditions near Bob that are retarded to the time $t-D/c$; similarly, Bob will calculate $\psi_1(x_1,t)$, but using boundary conditions near  Alice also retarded by $D/c$.  We will refer to these as ``retarded wavefunctions'' as a shorthand for ``contemporary wavefunctions based on retarded boundary conditions'', and write them as
\begin{eqnarray}
\Psi_A(x_1,x_2,t) &=& \psi_1(x_1,t)\psi_2(x_2,t-D/c)-\psi_1(x_2,t-D/c)\psi_2(x_1,t) \ , \nonumber \\
\Psi_B(x_1,x_2,t) &=& \psi_1(x_1,t-D/c)\psi_2(x_2,t)-\psi_1(x_2,t)\psi_2(x_1,t-D/c) \ . \notag
\end{eqnarray}

If the boundary conditions remain static, the wavefunctions evaluated with time-retarded boundary conditions become identical to ones with instantaneous boundary conditions, which means that Alice and Bob will then generate exactly the entangled results (that violate Bell's inequality).  If the boundary conditions do change, then it will take time for the new ``entanglement wave'' to propagate from Alice to Bob (or {\it vice versa}), and Bob, using ``old'' information, will calculate a trajectory different from one using non-local instantaneous information (we assume averages of such events will satisfy Bell's inequality and give a concrete illustration of this in the section below on trajectories for massive singlets).

\subsection{Nonlocality involving violation of Bell's Inequality}

We~\cite{mrc3,mrc4} and others~\cite{matzkin07} have shown that Bell's derivation of Bell's inequality,
  \begin{eqnarray}
|\langle AB\rangle - \langle A'B'\rangle| + |\langle AB'\rangle + \langle A'B\rangle|& \le & 2 \ , \label{eq:bellsymlimit}
\end{eqnarray}
 depended on a particular form of counterfactual definiteness -- that two measurement {\em results} can be considered to exist at the same time -- {\it  e.g.} that position ($A$) and momentum ($A'$) can be measured simultaneously -- something that is forbidden by the Heisenberg Uncertainty Principle.\footnote{This sneaks in when one tries to avoid ``contextuality'' without making a distinction between distant contexts (Bob's settings affecting Alice's results) and nearby ones (Alice not being able to make both settings simultaneously).} 
 
The ``instruction set'' paradigm of hidden variables~\cite{Mermin85} can also be viewed this way: ``the hidden variables are the results not measured'' -- not something that might be a proximate {\em cause} of a measurement (like a spin vector with some horizontal as well as vertical component, or what instruction sets {\em ought} to be), but the measurement result {\em itself}, since the instruction is plugged into the formula where the measured results are supposed to go.  But what can a measurement result be in the absence of a measurement -- the particle  going left (unnoticed, in a virtual horizontal magnet?) while we detect that it is going up in our vertically oriented magnet?  Clearly, if we cannot physically make the two measurements at the same time, we should not talk about the measurement results existing at the same time, even though such classical habits are hard to break.
 
   Avoidance of such counterfactual reasoning results in a weaker inequality,
 \begin{eqnarray}
|\langle AB\rangle - \langle A'B'\rangle| + |\langle AB'\rangle + \langle A'B\rangle|
   &     \le   2   + \min  \left( \right. \ &  \left| \langle B_1  B'_3 - B'_2 B_4  \rangle\right|  ,  \notag \\
                                               &  &  \left.  \left| \langle  A_1 A'_4 - A'_2  A_3\rangle \right| \ \right)  \ , \label{eq:gallupsqlimit}  
\end{eqnarray} 
consistent with the fact that one measurement can only be made strictly before or after the other.  This inequality also holds for any classical  system for which two measurements cannot be made simultaneously.\footnote{For example, even if a person could understand two questions that were asked of him simultaneously, there is no way that he could answer them simultaneously.  Such psychological ``measurements'' ({\it i.e.},``trick questions'') can be   incompatible, such that different answers result depending on the order the questions are asked, and  could require modeling with non-commutative models.} 
  Of course, this should not be a surprise, since one can easily show~\cite{R03,dBMR99} that the quantum-mechanical measurement operators (for 2-particle states) obey a similar inequality: 
 \begin{eqnarray}
\hat{S}_{Bell}^2 & = &  4\hat{I}  - [\hat{A},\hat{A'}] [\hat{B},\hat{B'}] \ . \label{eq:bellsquare}
\end{eqnarray}

A number of philosophical questions arise when we contemplate this relation:  since there is no factor of $\hbar$ multiplying the commutator terms on the right hand side of \Eq~\ref{eq:bellsquare}, how is it that a putatively nonlocal quantum mechanics can reduce to a local classical mechanics.  Furthermore, why is it only non-commutative experimental measurements that generate nonlocal results -- if Alice switches from $a$ to $a+180^o$ ($A \rightarrow -A$), then even if Bob continues to switch between $b$ and $b'$, nonlocality will not show up -- what kind of nonlocality only shows up when measured at some angles but not others?  The issues vanish if the explanation is that measurements are  contextual, as shown, for example, by Christian's model~\cite{C0703179}.  As we are about to see, the issues will also vanish if Bell's assumption that Alice and Bob's measurements, made a distance $D$ apart, can't be written as $A(a,\lambda)$ or $B(b,\lambda)$, but must be written as $A(a,b(t-D/c),\lambda)$ and $B(a(t-D/c), b, \lambda)$.  All derivations of Bell inequalities begin by assuming the $A(a,\lambda)$ form as the mathematical form of the locality postulate, but this is strictly true only if $A(a(t),\lambda)$ has no dependence on $b(t)$; a dependence on $b(t-D/c)$, while local, is causally  ``contextual''.  It is not clear if EPR would reject that out of hand given the alternative of true nonlocality.

\section{\dbb~Trajectories for Massive Singlets\label{secn:dbblaws}}

The \dbb~analog of waiting for a sound-wave to reach the second mass point in a Hooke-law spring is to be understood, not by retarding the second particle's position, but by using the appropriately retarded wavefunctions in the velocity law (as we have argued in section~\ref{secn:entangl}).  Imagine that Alice doesn't change her magnet, but Bob does.  In that case, the wavefunction inside Bob's magnetic volume will change (adiabatically, we presume) while his magnet is being switched, and a Dirac wave (a wave in the electron aether distinct from any electro-magnetic Maxwell wave) will propagate out from that region at  the speed of light toward Alice.  As long as Alice uses  retarded boundary conditions to calculate Bob's wavefunction, she can calculate both $x_1$ and $x_2$ in a causal manner for use in her velocity prescription for $\dot{x}_1$; similarly, if Bob uses his instantaneous wavefunction and  retarded boundary conditions for Alice's wavefunction, his $\dot{x}_2$ will be local and causal.

Durt and Pierseaux\cite{DP02} defined an EPR experiment based, presumably, on  typical Stern-Gerlach experiments.   For their case of two silver atoms in a singlet spin state, they assumed that the velocity of the atom was $10^4$ cm/s, and if we further assume that the overall size of the system is $\mathcal{O}$(100 cm), then Alice will know about Bob's new wavefunction in $\mathcal{O}(10^{-8}$ s), while the particles take $\mathcal{O}(10^{-2}$ s) to get from source to detector.  Unless ultra-relativistic particles are used, it would seem that there is no way for Alice {\em not} to know Bob's magnetic field setting in time to calculate a local expression  for her particle. 

{\em The locality loophole will be very hard to close for massive particles,  and \dbb~velocity prescriptions using wavefunctions based on retarded boundary conditions will provide a local hidden variable theory that violates Bell's inequality in perfect agreement with the quantum results.}

Holland~\cite[p. 469]{holland} and Durt \& Pierseaux\cite{DP02} have calculated the velocities of the two particles (using D\&P's notation, which distinguishes Left from Right, but for which we can read Alice and Bob; with the local $z$ axes parallel to the local magnetic fields, $\hat{z}_{L|R} \parallel \hat{B}_{L|R}$):
\begin{subequations}\label{eq:dprules}\begin{align}
\frac{dz_L}{dt} = \frac{k^2}{1+k^2t^2} t z_L +&
                                            \frac{s^2 e^{\left(\frac{\beta t^2}{1+k^2t^2}\frac{(z_L+z_R)}{2}\right)}
                                                      +c^2 e^{\left(\frac{\beta t^2}{1+k^2t^2}\frac{(z_L-z_R)}{2}\right)}
                                                      -c^2 e^{\left(\frac{\beta t^2}{1+k^2t^2}\frac{(-z_L+z_R)}{2}\right)}
                                                      -s^2 e^{\left(\frac{\beta t^2}{1+k^2t^2}\frac{(-z_L-z_R)}{2}\right)}   }
                                                     {s^2  e^{\left(\frac{\beta t^2}{1+k^2t^2}\frac{(z_L+z_R)}{2}\right)}
                                                      +c^2e^{\left(\frac{\beta t^2}{1+k^2t^2}\frac{(z_L-z_R)}{2}\right)}
                                                     +c^2e^{\left(\frac{\beta t^2}{1+k^2t^2}\frac{(-z_L+z_R)}{2}\right)}
                                                     +s^2 e^{\left(\frac{\beta t^2}{1+k^2t^2}\frac{(-z_L-z_R)}{2}\right)}  }  \notag \\
                                   &           \cdot \alpha t\left[ 2 - \frac{k^2 t^2}{1+k^2t^2}\right] \ ,  \label{eq:dprulesa}  \\
\frac{dz_R}{dt} = \frac{k^2}{1+k^2t^2} t z_R +&
                                            \frac{s^2 e^{\left(\frac{\beta t^2}{1+k^2t^2}\frac{(z_L+z_R)}{2}\right)}
                                                      -c^2 e^{\left(\frac{\beta t^2}{1+k^2t^2}\frac{(z_L-z_R)}{2}\right)}
                                                      +c^2 e^{\left(\frac{\beta t^2}{1+k^2t^2}\frac{(-z_L+z_R)}{2}\right)}
                                                      -s^2 e^{\left(\frac{\beta t^2}{1+k^2t^2}\frac{(-z_L-z_R)}{2}\right)}   }
                                                     {s^2  e^{\left(\frac{\beta t^2}{1+k^2t^2}\frac{(z_L+z_R)}{2}\right)}
                                                      +c^2e^{\left(\frac{\beta t^2}{1+k^2t^2}\frac{(z_L-z_R)}{2}\right)}
                                                     +c^2e^{\left(\frac{\beta t^2}{1+k^2t^2}\frac{(-z_L+z_R)}{2}\right)}
                                                     +s^2 e^{\left(\frac{\beta t^2}{1+k^2t^2}\frac{(-z_L-z_R)}{2}\right)}  } \notag \\
                                   &           \cdot\alpha t\left[ 2 - \frac{k^2 t^2}{1+k^2t^2}\right] \ ,  \label{eq:dprulesb}
 \end{align}\end{subequations}
 where 
 \begin{subequations}\begin{align}
 \alpha &=  \frac{dB}{dz} \mu/2m \ , \label{eq:dp0} \\
 \beta &= 2\alpha/ \delta r_0^2 \ , \\
 k &=  \hbar/2 m \delta r_0^2 \ , \\
 c  &\equiv  \cos(\theta_A - \theta_B)/2 \ \ , \ \ s\equiv \sin(\theta_A - \theta_B)/2 \ , 
 \end{align}\end{subequations}
 with $\mu$ is the magnetic moment, $m$ the mass of the silver atom, $\delta r_0$ the initial width of the Gaussian wavepacket and $dB/dz$ the magnetic field gradient.  $\theta_A$ and $\theta_B$ are the orientations of the magnetic fields with respect to a third, fixed, frame.   When $s\ne0$, the two $z$-axes are \emph{not} aligned and the arguments of the exponentials are \emph{not} the z-components of the relative and center-of-mass coordinates but a mixture of the $z$ and $x$ coordinates (assuming the particles separate from the source in the $\pm\hat{y}$ direction).
 
 \subsubsection{Local {\it vs.} Nonlocal Velocity Prescriptions}
 
For Alice to retard Bob's boundary condition in her instantiation of the velocity expressions means that she must use a causal value for $\theta_B$, and similarly, Bob must use only values for $\theta_A$ that he can know causally.  Each then uses both equations to model both trajectories, using only the (jointly known) initial location of the particles.

For a nonlocal prescription, Alice is free to use the instantaneous value of Bob's magnet setting.  Whether Alice continues to calculate both trajectories, or just ``knows'' Bob's instantaneous result is moot, since the two methods will result in the same answer. 

Recall that for massive particles, it is almost certain that the Dirac wave from Bob's changing magnetic field will overtake the Alice-bound particle while it is in the field-free region between the source and Alice's magnet/detector.  In that region, entangled Schr\"odinger wavefunctions generate the same simple velocity law as unentangled wavefunctions, $\vec{v}^A = -\vec{v}^B = -\hbar \vec{k}/m$.  Thus, we hypothesize that the changes in the Dirac wave will not affect the particle as it passes over it, and that when the particle finally reaches Alice, she will know the correct magnetic field setting to use for Bob's boundary condition and therefore impose the correct acceleration on her particle.  This local result of Alice's will generate a result identical to  the ``nonlocal'', Bell-violating result.

 \subsubsection{The Locality of \dbb~Trajectories for Massive Singlets in Aligned Fields}
 
 When Alice and Bob's detectors are both at the same angle, $\theta_A=\theta_B$, then $s=0, c=1$ and the velocity law, like Hooke's law, reduces to a function of the relative distance coordinate only:
 \begin{eqnarray}
 \frac{d(z_L-z_R)}{dt} &=& \frac{k^2}{1+k^2t^2} t (z_L-z_R)  + \alpha t\left[ 2 - \frac{k^2 t^2}{1+k^2t^2}\right] \cdot \tanh{\left(\frac{\beta t^2}{1+k^2t^2}\frac{(z_L-z_R)}{2}\right)} \ . \label{eq:dpmatch}
 \end{eqnarray}

 Since we have shown that the Hooke (or Coulomb) type force law is not {\em actually} nonlocal, being an idealization of a more fundamental local continuum model, we can conclude that the entangled particle velocity law -- at least when encountering aligned magnets -- is also generating {\em local} trajectories, as long as the system is isolated.  This is important because a recent paper~\cite{Norsen06a} purports to show that ``Bell Locality'' imposes certain constraints on \emph{any} hidden variable theory, such that it must then obey Bell's inequality.  Since we have just shown that the \dbb~interpretation becomes a local model (and satisfies the constraints Norsen proposed for the case of aligned magnet settings), we have an obvious contradiction since we know that our interpretation is local and  the \dbb~model will reproduce the quantum mechanical results. The contradiction can be resolved by realizing that Bell's inequality implicitly requires  hidden variable theories to violate Heisenberg's uncertainty principle~\cite{mrc3} -- whereas \dbb~can predict counterfactual results ({\it i.e.}, $A(a,\lambda, t)$ and $A(a',\lambda, t')$), but it can't predict them simultaneously, but only $A(\frac{a+a'}{2}, \lambda,t)$ in such a case.

 The fact that Alice's velocity depends  on $c$ and $s$ (in particular, on $\theta_B$), makes this \dbb~trajectory violate one of Bell's assumptions,\footnote{Examining Wigner's derivation of Bell's inequality\cite{Wig76}, it is clear that local time-delayed knowledge of one's partner's settings is explicitly excluded in his proof  as well.} 
 that the measurement result, $A$, be written as $A(a,\lambda)$, and not as $A(a,b,\lambda)$, (or $A(a(t),b(t-\tau),\lambda)$).  In the case that a Dirac wave has time to propagate from Bob to Alice, then there is no actual nonlocality, just a  nonlinear sensitivity to (remote) initial conditions that is only revealed by entangled wavefunctions -- indeed all of the physical phenomena that tout nonlocality only require entanglement in order to be understood, and we might refer to this as ``contextuality at a distance''.

 \subsubsection{Computer Experiments}
 
 We have programmed an EPR experiment, using the Durt and Pierseaux velocity prescription to determine the motion of neutral silver atoms in a 30 cm magnet with a field gradient of $10^4$ G/cm, and initial spread in the Gaussian wavefunction of $10^{-3}$ cm, so that $\alpha = 2.58\cdot10^5$ cm s$^{-2}$, $\beta= 5.17\cdot10^{11}$ cm$^{-1}$s$^{-2}$, and $k=2.94$.  The values in table~\ref{tab:dbbb} resulted from the tracking of 4000 pairs of particles, where new magnet settings were randomly chosen as each pair was launched (this means that even though the particles took a millisecond to traverse the magnet, and presumably as long to get to the magnet, we pretend that the speed of light is even slower, so that Alice and Bob are spacelike separated). 
 
 Starting with the bottom two rows of the table, if Alice and Bob use non-local knowledge of the other's instantaneous magnet setting, and if there are no physics-induced detector inefficiencies,  the coincidence rate will be the same as the singles rate in each detector, so that normalizing by singles (or coincidences) will result in a small standard deviation (0.036, third row); if there are physics-induced inefficiencies (see section~\ref{secn:ineff} below), then the number of coincidences will be about 1/4 the number of single events, leading to a standard deviation twice as large (bottom row), in both cases violating Bell's inequality.
 
 If Alice and Bob only use local knowledge of the other's setting ({\it i.e.}, retarded boundary conditions), and if there are no detector inefficiencies, then they will calculate the correct position of their partner's particle only half the time (on average) and the Bell parameter they measure will be less than 2 (1.3, but with a small standard deviation, first row).  If there are processes that throw particles out of the beamline before getting to a magnet that has just been changed, then Alice and Bob will find their coincidence rate to be half their singles rate, but by normalizing to the  coincidences, they will calculate a Bell parameter of 2.8 with a larger standard deviation (0.07, second row). 
 
  If Alice and Bob change their magnets without affecting the oncoming particles, and if the particles move at subluminal speeds ($v/c \sim 10^{-6}$) so that the Dirac wave outruns the particle, then Alice and Bob will be using the ``effectively instantaneous'' values of each others magnet settings. In such a case, they will measure a causal (local) violation of Bell's inequality with a small standard deviation (identical to results in the third row). 
  
 \begin{table}
\begin{center}
\begin{tabular}{rcccc}
   Locality  &  Normalization  & Detector Efficiency \ \ \ &  \ {$ S_{Bell}$} \ &  \ \ \ \ {std. dev.} \  \\ \hline
Loc  & singles           & Efficient           &-1.31946 &$ \pm$  .03652 \\
Loc   & coincidences  & Inefficient           &-2.76893 & $ \pm$  .07086   \\
NonL & singles         &  Efficient         &-2.77554 & $\pm$  .03652   \\
 NonL & coincidences  & Inefficient       &-2.76893 & $\pm$  .07086 \\ \hline
\end{tabular}
\end{center}
\caption{Calculationally measured Bell parameters for various ``computational'' configurations.} \label{tab:dbbb}
\end{table}

\section{Physical But Unfair Sampling Effects\label{secn:ineff}}

In order to experimentally establish nonlocality with massive particles, it is necessary to  switch the magnet settings on Alice and Bob's detectors in a random manner, and at a rate greater than the source emission rate.  Clearly if two or more singlets are emitted before either detector changes settings, there will be a much higher percentage of 	``locally'' describable events than the potentially non-local, for which Alice and Bob, using the old (wrong) settings, will predict the wrong event behavior.  On the other hand, the settings should not change so frequently as to change while the particle is inside the magnet, nor should they change and change back again before the particle gets to the detector.

Using the Durt and Pierseaux expressions with the time-lagged values of the other's magnet setting, we have shown, {\it via} computer experiments, that the Bell parameter drops to a value below 2.0  when Alice and Bob randomly pick a new detector setting after each singlet is launched (compared to $2\sqrt{2}$ when using the instantaneous  values of both).  If such an experiment could actually be performed, we would expect to see experiments measure values significantly less than 2 when (if) the ``locality loophole'' is closed, but $2\sqrt{2}$ if it is not.

However, we remind the reader that in such an experiment~\cite{GG05}, the magnetic field will be of order 1-10 T or $\sim10^5$ G (although Durt and Pierseaux never mention the field strength), with spatial gradients of the field of order $10^2$ T/m ($10^4$ G/cm).  The magnet can be switched either mechanically, or by the equivalent of energizing two (opposite) poles of a quadrupole while de-energizing the other two poles.   This change of magnetic field will result in the emission of a real electromagnetic (Maxwell) wave travelling down the flight-path toward the source of the particles (as well as in all other directions), and the wave will have a magnetic field in the direction of $a - a'$ (or $b - b'$).  In other words, if Alice's settings for her magnetic field ($\hat{B}$) are $a=0$ and $a=\pi/2$, then her $\Delta\hat{B}$ will point at $-\pi/4$ or $+3\pi/4$. 

For the sake of clarity, let us assume that on all relevant timescales for this problem, it takes no time to switch the magnet from one setting to another.  Thus Bob's full $\Delta B$ leaves the front end of his magnet and travels toward Alice, and $30/c$ seconds later (assuming that the magnet is $L=30$ cm long), $\Delta B$ drops back to zero as the signal from the back end of the magnet leaves the front on its way to Alice.  Within this virtual magnet volume that is sweeping from Bob to Alice,  the gradients of the magnetic field are the same size as in the magnet --  $\frac{dB_z}{dz} = 10^4$ G/cm.\footnote{ Although Durt and Pierseaux assume for simplicity of exposition that no other component of $\vec{B}$ has any gradients, this is not physical -- in order for $\nabla\cdot\vec{B}=0$, it must be the case that at least $\frac{dB_x}{dx} = - \frac{dB_z}{dz}$ if there is no gradient in the direction of motion,  but more generally, $\frac{dB_x}{dx}\approx \frac{dB_y}{dy}\approx -\frac{1}{2}\frac{dB_z}{dz}$.  In this case, as long as \emph{any} electromagnetic field is traveling, the magnitude of the spatial gradients in any direction will be comparable to any other direction.} 
(One might argue that electromagnetic fields should fall off as $r^{-2}$, making the magnitude of the field too small to have any effect.  This would be true in free space, but for a neutral massive particle (\eg,~Ag) to cover a distance of order meters from source to detector, it will need to travel in an evacuated beamline, and those are usually made of steel; to the extent that this will act as a wave guide, the $\Delta {B}$ field strength will not attenuate.)

Since the Durt and Pierseaux velocity formulae only depend on this derivative of the magnetic field, we can treat the EM wave washing over the particle in flight as if it were in a shorter magnet that would take $L/c= 10^{-9}$ s to traverse, and use \Eqs~\eqref{eq:dprules} to calculate that the particle will experience a transverse velocity {\em before} it even gets to the magnet, proportional to that shorter time.  The transverse velocity for our typical silver atom will be about a part per million of what it would have been after leaving the real magnet, since the time in the real magnet is $\mathcal{O}(10^{-3}$ s) compared to  the nanosecond  ``in'' the EM wave, resulting in an insignificant effect.  

Now consider a faster particle, say one moving $10^3$ times faster, so that it only spends  $\mathcal{O}(10^{-6}$ s) in the magnet (assuming for the sake of argument that the very small  deflections will still be resolvable).  The $30 \mbox{ cm}$ long $\Delta B$ wave will still pass over the particle in 1 ns, so the transverse velocity will now be a part per thousand of those making the resolvable deflections.  This too will probably not result in any lost particles. 

However, if we now go for the ultimate neutral fermion, and send entangled pairs of neutrino's at the Stern-Gerlach apparatus, the neutrino will spend only twice as much time in the magnet as under the influence of the $\Delta B$ wave.  This means the transverse velocity experienced between the source and the detector will be half that it would experience in the detector, and the deflection a quarter.  Thus, for light particles moving at near light-speeds,  one can expect that the $\Delta B$  field will kick any particle in flight entirely out of the experiment,  deflecting it so much that it can no longer fit through any collimators at the entrance to the magnet.\footnote{The fact that a neutrino's mass is $\mathcal{O}$(1) eV compared to $\mathcal{O}(10^{11})$ eV for Ag$^{108}$ means that $\alpha$ in \Eq~\eqref{eq:dp0} will increase a lot more than the time in the magnet will decrease -- $(\alpha \Delta t)_{\nu} \sim 10^5 (\alpha \Delta t)_{Ag}$ -- assuming the magnetic moments are comparable; alternatively, while a silver atom at $v/c=0.99$ would require a magnet thousands of kilometers long to generate the same deflections if the mass in the expression for $\alpha$  (\Eq~\eqref{eq:dp0}) goes over to $\gamma m_0$ (so that $\alpha$ drops by a factor  $\mathcal{O}(10^{-2})$ and $\Delta t$ drops by $\mathcal{O}(10^{-6})$), it would avoid being kicked out of the beamline before the detector by a changing field.} 
Unfortunately, neutrino's have their own detector efficiency problems.

The simplest experimental signal of such a phenomenon would be to make absolute count-rate measurements, so that if Alice and Bob make no changes to their magnets, one measures a quiescent singles count rate, $Q_1$ somewhat larger than the quiescent councidence rate $C_2$ (the latter being dependent on the square of the efficiency of nearly perfect detectors).  If Alice and Bob now randomly switch their magnets once on average during the source-detector flight time, in the absence of this detector inefficiency, their coincidence rate should remain $C_2$;  but if this effect is present, then the coincidence rate will drop to $C_2'\sim C_2/4$,  the singles rate to $Q_1'\sim Q_1/2$, and these relative drops should be the same regardless of the efficiency of the individual detectors.

\section{Photon EPR experiments}

Tittel {\it et al.}~\cite{TB98,TBZG98} performed an EPR experiment in Geneva that atempted to close the locality loophole by arranging that the photon in one arm of the experiment went through a beam-splitter into  one of  two analyzers.  As the authors  point out, a hidden variable could determine into  which one of the two it would go, preserving locality, since the setup was static.  The authors dismiss this, arguing that a quantum-mechanical random number generator (the beam splitter) is the best that can be imagined.  Given that we have argued that entanglement is the underlying causal explanation for the violations of Bell's inequalities, and entanglement relies on the quantum (Dirac) fields that permeate space, it requires no additional stretch of the imagination to believe that the \dbb-ish behavior  of the photons at the beam splitter is also determined by similar (Maxwell) fields that determine the Bell violations in the detectors themselves.  We conclude that this experiment has a gaping locality loophole.

The Innsbruck experiment of Weihs {\it  et al.}~\cite{WZ98} was designed to close the locality loophole, and did so if the assumption of fair  sampling is valid.  In this experiment, electro-optical modulators were used to rotate the polarization of the photon before it entered the polarizing beam-splitter/detector.  These modulators are in some sense the analog of the magnetic field in the Stern-Gerlach apparatus, although no deviation of photon trajectories occurs inside the devices.  If a detector loophole  were the only flaw in this experiment, then one would have to rely  on some property of the modulators, that, for example when switched, induces a loss of total-internal-reflection in the fiber-optic cable, so that photons are lost from the fiber if the switch is made.  Although it is highly unlikely that such an effect would have gone unnoticed, it would be interesting to know if the coincidence counting rates vary when the modulators are turned off during the course of a setting measurement ({\it e.g.}, $\langle AB'\rangle$), compared to when the modulators switch between each coincidence event. 

Given the nature of our local \dbb~trajectories for massive particles, there is no conceivable way such a model could account for the spacelike separated EPR results without a detector loophole.  On the other hand, there is no \dbb~prescription for photons~\cite[p.  541]{holland}, and it may be that the Clifford-algebra model of Christian~\cite{C0703179} may be a more correct idealization in this context than our fermion model, in which case the Innsbruck data may also be explainable by a local model.

\section{Conclusion}

\subsection{The Meaning of Nonlocality}

The controversy over the EPR paradox has depended on ambiguous concepts of locality and contextuality.  EPR were wrong (and Bohr was right as our \dbb~expressions show) in that Alice's measurement result {\em can} depend on the measurement that Bob makes, if the two particles are entangled.  EPR would have been right to argue that if there is such a distance effect, it must be time-retarded in order to be causal (we thus predict that  perfect photon detectors will show obedience to Bell's inequality in Innsbruck-type experiments):  time-retarded ``action at a distance'' is {\em not} nonlocal, but it is contextual.\footnote{EPR assumed that either QM was incomplete, or that the physical quantities corresponding to non-commuting operators weren't simultaneously real; they did not distinguish~\cite[p. 780]{EPR35} between the measurement result's reality -- the positive or negative deflections in a magnet (``...we must consider the quantity $P$ [{\it i.e.}, $\sigma_x=\pm 1$] as being an element of reality...'')  -- and the ``hidden'' properties of a particle that caused such deflections, even though they referred earlier~\cite[p. 777]{EPR35} to ``...an element of physical reality {\em corresponding} to this physical quantity'' [my emphases].  A \dbb~interpretation of Bohm's spin example would distinguish between the left or right value of the hidden position coordinate, $x$, which results in $\sigma_x =\pm 1$, and the up or down value of the vertical position coordinate, $y$, that results in $\sigma_y=\pm 1$. A magnetic field cannot be devised that measures both deflections simultaneously,  but both hidden positions exist (simultaneously) even if only one is relevant to a given measurement.} 

Because EPR rejected contextuality at a distance, and equated that rejection to locality, Bell and others' derivations of inequalities, as part of mathmatically translating  ``no context at a distance'' into $A(a,\lambda)$ and $B(b,\lambda)$, have implicitly assumed that there can be no ``local contextuality'' either, so that even if $A(a,\lambda)$ is measured, $A(a', \lambda)$ is assumed to exist, and is used in formulae in a manner to suggest that it {\em could have been measured} at the same time and place, thereby limiting their inequality to hidden variable theories that violate Heisenberg's Uncertainty principle.  If hidden variable theories are restricted by assumption to modeling commuting operators, it is not surprising that they fail to model non-commuting ones;  Christian's Clifford algebra~\cite{C0703179} provides a counterexample of a local model with non-commuting properties.

\subsection{R\'esum\'e}

 The ``manifestly nonlocal'' equations of motion, be they Hooke's or \dbb's, are valid  only as long as the system they are intended to model is isolated, which means only as long as no one {\em tries} to signal by grabbing one of the no-longer-isolated parts. Any attempt to change the boundary conditions destroys the assumption of isolation, invalidates the equation(s), and moots the point of whether the equation can signal or not.  We have also seen that two-body idealizations are approximations of an underlying local continuum model; Hooke has his Navier-Stokes, but there is still work to bring QED/QCD to an equally satisfactory footing to underlie the \dbb~interpretation of Schr\"odinger/Dirac.

By analogy with Hooke's Law, we have developed a local interpretation of entangled trajectories given time independent wavefunctions.  We have also shown that this interpretation, for time dependent wavefunctions, can be made truly nonlocal by using, unjustifiably,  non-retarded values of the wavefunctions.  This makes the locus of \dbb~nonlocality reside in superluminal Dirac waves, which is patently absurd. On the other hand, we can calculate local, causal trajectories by using appropriately time-retarded wavefunction(s) in the ``calculation'' of the velocities for these particles, making this the first local model (to our knowledge) that is physics-based and not artificially contrived. Noting that the Dirac waves for massive particles are such that any changes in those waves (and their entanglement) can only move at the speed of light  while the massive particles guided by them  move at subluminal velocities,  we have concluded that it will be very difficult to close the locality loophole for such particles~\cite{Rowe01}, but that if it is closed, we predict they will satisfy Bell's inequality.

Finally, we have also shown that act of changing the magnetic fields generates an EM wave that can affect the oncoming particles and could, for relativistic particles, kick them out of the detector, thereby creating a detector loophole  in exactly the conditions that would otherwise close the locality loophole.


\bibliography{dbb2}

\begin{thebibliography}{31}
\expandafter\ifx\csname natexlab\endcsname\relax\def\natexlab#1{#1}\fi
\expandafter\ifx\csname bibnamefont\endcsname\relax
  \def\bibnamefont#1{#1}\fi
\expandafter\ifx\csname bibfnamefont\endcsname\relax
  \def\bibfnamefont#1{#1}\fi
\expandafter\ifx\csname citenamefont\endcsname\relax
  \def\citenamefont#1{#1}\fi
\expandafter\ifx\csname url\endcsname\relax
  \def\url#1{\texttt{#1}}\fi
\expandafter\ifx\csname urlprefix\endcsname\relax\def\urlprefix{URL }\fi
\providecommand{\bibinfo}[2]{#2}
\providecommand{\eprint}[2][]{\url{#2}}

\bibitem[{\citenamefont{Clover}(2003)}]{mrc1}
\bibinfo{author}{\bibfnamefont{M.}~\bibnamefont{Clover}},
  \emph{\bibinfo{title}{The {I}nnsbruck {EPR} experiment: A time-retarded local
  description of space-like separated correlations}} (\bibinfo{year}{2003}),
  \eprint{quant-ph/0304115v2}.

\bibitem[{\citenamefont{Philipp and Hess}(2002)}]{PH02}
\bibinfo{author}{\bibfnamefont{W.}~\bibnamefont{Philipp}} \bibnamefont{and}
  \bibinfo{author}{\bibfnamefont{K.}~\bibnamefont{Hess}},
  \emph{\bibinfo{title}{A local mathematical model for {EPR} experiments}}
  (\bibinfo{year}{2002}), \eprint{quant-ph/0212085}.

\bibitem[{\citenamefont{Thompson and Holstein}(1996)}]{TH02}
\bibinfo{author}{\bibfnamefont{C.~H.} \bibnamefont{Thompson}} \bibnamefont{and}
  \bibinfo{author}{\bibfnamefont{H.}~\bibnamefont{Holstein}},
  \bibinfo{journal}{Found.\ Phys. \ Lett.} \textbf{\bibinfo{volume}{9}},
  \bibinfo{pages}{357} (\bibinfo{year}{1996}).

\bibitem[{\citenamefont{Gisin and Zbinden}(1999)}]{GZ99}
\bibinfo{author}{\bibfnamefont{N.}~\bibnamefont{Gisin}} \bibnamefont{and}
  \bibinfo{author}{\bibfnamefont{H.}~\bibnamefont{Zbinden}},
  \emph{\bibinfo{title}{Bell inequality and the locality loophole: {A}ctive
  versus passive switches}} (\bibinfo{year}{1999}).

\bibitem[{\citenamefont{Gisin and Gisin}(1999)}]{GG99}
\bibinfo{author}{\bibfnamefont{N.}~\bibnamefont{Gisin}} \bibnamefont{and}
  \bibinfo{author}{\bibfnamefont{B.}~\bibnamefont{Gisin}},
  \emph{\bibinfo{title}{A local hidden variable model of quantum correlation
  exploiting the dectection loophole}} (\bibinfo{year}{1999}),
  \eprint{quant-ph/9905018}.

\bibitem[{\citenamefont{Szabo and Fine}(2002)}]{SF02}
\bibinfo{author}{\bibfnamefont{L.~E.} \bibnamefont{Szabo}} \bibnamefont{and}
  \bibinfo{author}{\bibfnamefont{A.}~\bibnamefont{Fine}},
  \bibinfo{journal}{Phys.\ Lett.\ A} \textbf{\bibinfo{volume}{295}},
  \bibinfo{pages}{229} (\bibinfo{year}{2002}), \eprint{quant-ph/0007102}.

\bibitem[{\citenamefont{Larsson}(1999)}]{L99}
\bibinfo{author}{\bibfnamefont{J.}~\bibnamefont{Larsson}},
  \bibinfo{journal}{Phys.\ Lett. \ A} \textbf{\bibinfo{volume}{256}},
  \bibinfo{pages}{245} (\bibinfo{year}{1999}).

\bibitem[{\citenamefont{Leggett}(2003)}]{L03}
\bibinfo{author}{\bibfnamefont{A.~J.} \bibnamefont{Leggett}},
  \bibinfo{journal}{Found.\ Phys.} \textbf{\bibinfo{volume}{33}},
  \bibinfo{pages}{1469} (\bibinfo{year}{2003}).

\bibitem[{\citenamefont{Accardi and Khrennikov}(2006)}]{Ac06}
\bibinfo{author}{\bibfnamefont{L.}~\bibnamefont{Accardi}} \bibnamefont{and}
  \bibinfo{author}{\bibfnamefont{A.}~\bibnamefont{Khrennikov}},
  \emph{\bibinfo{title}{Chameleon effect, the range of values hypothesis and
  reproducing the {EPR}-{B}ohm correlations}} (\bibinfo{year}{2006}),
  \eprint{quant-ph/0611259v1}.

\bibitem[{\citenamefont{Christian}(2007)}]{C0703179}
\bibinfo{author}{\bibfnamefont{J.}~\bibnamefont{Christian}},
  \emph{\bibinfo{title}{Disproof of {B}ell's {T}heorem by {C}lifford {A}lgebra
  valued {L}ocal {V}ariables}} (\bibinfo{year}{2007}),
  \eprint{quant-ph/0703179}.

\bibitem[{\citenamefont{Bohm}(1952)}]{B52a}
\bibinfo{author}{\bibfnamefont{D.}~\bibnamefont{Bohm}},
  \bibinfo{journal}{Phys.\ Rev.} \textbf{\bibinfo{volume}{85}},
  \bibinfo{pages}{180} (\bibinfo{year}{1952}).

\bibitem[{\citenamefont{Bell}(1993)}]{B66}
\bibinfo{author}{\bibfnamefont{J.}~\bibnamefont{Bell}}, in
  \emph{\bibinfo{booktitle}{Speakable and Unspeakable in Quantum Mechanics}}
  (\bibinfo{publisher}{Cambridge University Press}, \bibinfo{year}{1993}), pp.
  \bibinfo{pages}{1--13}.

\bibitem[{\citenamefont{Durt and Pierseaux}(2002)}]{DP02}
\bibinfo{author}{\bibfnamefont{T.}~\bibnamefont{Durt}} \bibnamefont{and}
  \bibinfo{author}{\bibfnamefont{Y.}~\bibnamefont{Pierseaux}},
  \bibinfo{journal}{Phys. \ Rev. \ A} \textbf{\bibinfo{volume}{66}},
  \bibinfo{pages}{052109} (\bibinfo{year}{2002}).

\bibitem[{\citenamefont{Holland}(1999)}]{H99}
\bibinfo{author}{\bibfnamefont{P.}~\bibnamefont{Holland}},
  \bibinfo{journal}{Phys.\ Rev.\ A} \textbf{\bibinfo{volume}{60}},
  \bibinfo{pages}{4326} (\bibinfo{year}{1999}).

\bibitem[{\citenamefont{Squires}(1995)}]{Sq95}
\bibinfo{author}{\bibfnamefont{E.}~\bibnamefont{Squires}},
  \emph{\bibinfo{title}{Lorentz invariant {B}ohmian mechanics}}
  (\bibinfo{year}{1995}), \eprint{quant-ph/9508014v2}.

\bibitem[{\citenamefont{Holland}(1993)}]{holland}
\bibinfo{author}{\bibfnamefont{P.~R.} \bibnamefont{Holland}},
  \emph{\bibinfo{title}{The Quantum Theory of Motion, An Account of the de
  Broglie-Bohm Causal Interpretation of Quantum Mechanics}}
  (\bibinfo{publisher}{Cambridge University Press}, \bibinfo{year}{1993}).

\bibitem[{\citenamefont{Clover}(2004)}]{mrc3}
\bibinfo{author}{\bibfnamefont{M.}~\bibnamefont{Clover}},
  \emph{\bibinfo{title}{Bell's {T}heorem: A new {D}erivation that preserves
  {H}eisenberg and {L}ocality}} (\bibinfo{year}{2004}),
  \eprint{quant-ph/0409058v2}.

\bibitem[{\citenamefont{Clover}(2005)}]{mrc4}
\bibinfo{author}{\bibfnamefont{M.}~\bibnamefont{Clover}},
  \emph{\bibinfo{title}{Bell's {T}heorem: A critique}} (\bibinfo{year}{2005}),
  \eprint{quant-ph/0502016}.

\bibitem[{\citenamefont{Mackman and Squires}(1995)}]{MS95}
\bibinfo{author}{\bibfnamefont{S.}~\bibnamefont{Mackman}} \bibnamefont{and}
  \bibinfo{author}{\bibfnamefont{E.}~\bibnamefont{Squires}},
  \bibinfo{journal}{Foundations of Physics} \textbf{\bibinfo{volume}{25}},
  \bibinfo{pages}{391} (\bibinfo{year}{1995}).

\bibitem[{\citenamefont{Rice}(1997)}]{Rice97}
\bibinfo{author}{\bibfnamefont{D.}~\bibnamefont{Rice}}, \bibinfo{journal}{Am.\
  J.\ Phys.} \textbf{\bibinfo{volume}{65}}, \bibinfo{pages}{144}
  (\bibinfo{year}{1997}).

\bibitem[{\citenamefont{Matzkin}(2007)}]{matzkin07}
\bibinfo{author}{\bibfnamefont{A.}~\bibnamefont{Matzkin}},
  \emph{\bibinfo{title}{Classical statistical distributions can violate
  {B}ell's inequalities}} (\bibinfo{year}{2007}), \eprint{quant-ph/0703251}.

\bibitem[{\citenamefont{Rizzi}(2003)}]{R03}
\bibinfo{author}{\bibfnamefont{A.}~\bibnamefont{Rizzi}},
  \emph{\bibinfo{title}{The {M}eaning of {B}ell's {T}heorem}}
  (\bibinfo{year}{2003}), \eprint{quant-ph/0310098v1}.

\bibitem[{\citenamefont{de~Baere et~al.}(1999)\citenamefont{de~Baere, Mann, and
  Revzen}}]{dBMR99}
\bibinfo{author}{\bibfnamefont{W.}~\bibnamefont{de~Baere}},
  \bibinfo{author}{\bibfnamefont{A.}~\bibnamefont{Mann}}, \bibnamefont{and}
  \bibinfo{author}{\bibfnamefont{M.}~\bibnamefont{Revzen}},
  \bibinfo{journal}{Found. \ Phys.} \textbf{\bibinfo{volume}{29}},
  \bibinfo{pages}{67} (\bibinfo{year}{1999}).

\bibitem[{\citenamefont{Norsen}(2006)}]{Norsen06a}
\bibinfo{author}{\bibfnamefont{T.}~\bibnamefont{Norsen}},
  \emph{\bibinfo{title}{Bell {L}ocality and the {N}onlocal {C}haracter of
  {N}ature}} (\bibinfo{year}{2006}), \eprint{quant-ph/0601205}.

\bibitem[{\citenamefont{Gondran and Gondran}(2005)}]{GG05}
\bibinfo{author}{\bibfnamefont{M.}~\bibnamefont{Gondran}} \bibnamefont{and}
  \bibinfo{author}{\bibfnamefont{A.}~\bibnamefont{Gondran}},
  \emph{\bibinfo{title}{A complete analysis of the {S}tern-{G}erlach experiment
  using {P}auli spinors}} (\bibinfo{year}{2005}), \eprint{quant-ph/0511276}.

\bibitem[{\citenamefont{Tittel et~al.}(1998{\natexlab{a}})\citenamefont{Tittel,
  Brendel, Gisin, Herzog, Zbinden, and Gisin}}]{TB98}
\bibinfo{author}{\bibfnamefont{W.}~\bibnamefont{Tittel}},
  \bibinfo{author}{\bibfnamefont{J.}~\bibnamefont{Brendel}},
  \bibinfo{author}{\bibfnamefont{B.}~\bibnamefont{Gisin}},
  \bibinfo{author}{\bibfnamefont{T.}~\bibnamefont{Herzog}},
  \bibinfo{author}{\bibfnamefont{H.}~\bibnamefont{Zbinden}}, \bibnamefont{and}
  \bibinfo{author}{\bibfnamefont{N.}~\bibnamefont{Gisin}},
  \bibinfo{journal}{Phys.\ Rev.\ A} \textbf{\bibinfo{volume}{57}},
  \bibinfo{pages}{3229} (\bibinfo{year}{1998}{\natexlab{a}}).

\bibitem[{\citenamefont{Tittel et~al.}(1998{\natexlab{b}})\citenamefont{Tittel,
  Brendel, Zbinden, and Gisin}}]{TBZG98}
\bibinfo{author}{\bibfnamefont{W.}~\bibnamefont{Tittel}},
  \bibinfo{author}{\bibfnamefont{J.}~\bibnamefont{Brendel}},
  \bibinfo{author}{\bibfnamefont{H.}~\bibnamefont{Zbinden}}, \bibnamefont{and}
  \bibinfo{author}{\bibfnamefont{N.}~\bibnamefont{Gisin}},
  \bibinfo{journal}{Phys.\ Rev.\ Lett.} \textbf{\bibinfo{volume}{81}},
  \bibinfo{pages}{3563} (\bibinfo{year}{1998}{\natexlab{b}}).

\bibitem[{\citenamefont{Weihs et~al.}(1998)\citenamefont{Weihs, Jennewein,
  Simon, Weinfurter, and Zeilinger}}]{WZ98}
\bibinfo{author}{\bibfnamefont{G.}~\bibnamefont{Weihs}},
  \bibinfo{author}{\bibfnamefont{T.}~\bibnamefont{Jennewein}},
  \bibinfo{author}{\bibfnamefont{C.}~\bibnamefont{Simon}},
  \bibinfo{author}{\bibfnamefont{H.}~\bibnamefont{Weinfurter}},
  \bibnamefont{and}
  \bibinfo{author}{\bibfnamefont{A.}~\bibnamefont{Zeilinger}},
  \bibinfo{journal}{Phys.\ Rev.\ Lett.} \textbf{\bibinfo{volume}{81}},
  \bibinfo{pages}{5039} (\bibinfo{year}{1998}).

\bibitem[{\citenamefont{Rowe et~al.}(2001)\citenamefont{Rowe, Kielpinski,
  Meyer, Sackett, Itano, Monroe, and Windeland}}]{Rowe01}
\bibinfo{author}{\bibfnamefont{M.~A.} \bibnamefont{Rowe}},
  \bibinfo{author}{\bibfnamefont{D.}~\bibnamefont{Kielpinski}},
  \bibinfo{author}{\bibfnamefont{V.}~\bibnamefont{Meyer}},
  \bibinfo{author}{\bibfnamefont{C.~A.} \bibnamefont{Sackett}},
  \bibinfo{author}{\bibfnamefont{W.~M.} \bibnamefont{Itano}},
  \bibinfo{author}{\bibfnamefont{C.}~\bibnamefont{Monroe}}, \bibnamefont{and}
  \bibinfo{author}{\bibfnamefont{D.~J.} \bibnamefont{Windeland}},
  \bibinfo{journal}{Nature} \textbf{\bibinfo{volume}{409}},
  \bibinfo{pages}{791} (\bibinfo{year}{2001}).

\bibitem[{\citenamefont{Wigner}(1983)}]{Wig76}
\bibinfo{author}{\bibfnamefont{E.~P.} \bibnamefont{Wigner}}, in
  \emph{\bibinfo{booktitle}{Quantum Theory and Measurement}}, edited by
  \bibinfo{editor}{\bibfnamefont{J.~A.} \bibnamefont{Wheeler}}
  \bibnamefont{and} \bibinfo{editor}{\bibfnamefont{W.~H.} \bibnamefont{Zurek}}
  (\bibinfo{publisher}{Princeton University Press}, \bibinfo{year}{1983}), p.
  \bibinfo{pages}{260}.

\bibitem[{\citenamefont{Einstein et~al.}(1935)\citenamefont{Einstein, Podolsky,
  and Rosen}}]{EPR35}
\bibinfo{author}{\bibfnamefont{A.}~\bibnamefont{Einstein}},
  \bibinfo{author}{\bibfnamefont{B.}~\bibnamefont{Podolsky}}, \bibnamefont{and}
  \bibinfo{author}{\bibfnamefont{N.}~\bibnamefont{Rosen}},
  \bibinfo{journal}{Phys.\ Rev.} \textbf{\bibinfo{volume}{47}},
  \bibinfo{pages}{777} (\bibinfo{year}{1935}).

\end{thebibliography}

\end{document}